\documentclass[useAMS,usenatbib,usegraphicx]{mn2e}
\bibpunct[]{(}{)}{;}{a}{,}{,}

 \usepackage[T1]{fontenc}
 \usepackage{ae,aecompl}

\title[Discovery of a nearby L-T transition object in the Southern Galactic plane
]{Discovery of a nearby L-T transition object in the Southern Galactic plane\thanks{Based on observations collected at the European Southern Observatory (ESO), Chile (NTT/SOFI program 077C.0117). Also data from the 2MASS project (University of Massachusetts and IPAC/Caltech, USA), and the SuperCOSMOS Sky Survey and Science Archive (SSS and SSA respectively)}}
\author[S. L. Folkes, D. J. Pinfield, T. R. Kendall and H. R. A. Jones]{S. L. Folkes,\thanks{E-mail:
s.l.folkes@herts.ac.uk} D. J. Pinfield, T. R. Kendall and H. R. A. Jones\\Centre for Astrophysics Research, Science and Technology  Research Institute, University of Hertfordshire, College Lane,\\Hatfiled, AL10 9AB, United Kingdom}

\newcommand{\JK}{{\em J-K}$_{s}$}
\newcommand{\HK}{{\em H-K}$_{s}$}

\newcommand{\JH}{{\em J-H}}
\newcommand{\Ks}{{\em K}$_{s}$}

\newcommand{\Rone}{{\em R}$_{63F}$}
\newcommand{\Rtwo}{{\em R}$_{59F}$}
\newcommand{\Ivn}{{\em I}$_{N}$}

\begin{document}

\date{Received ??; in original form 2006 ??}

\pagerange{\pageref{firstpage}--\pageref{lastpage}} \pubyear{2002}

\maketitle

\label{firstpage}
\bibliographystyle{mn2e}
\begin{abstract}
We present the discovery of 2MASS J11263991-5003550 identified as part of an ongoing survey to discover ultra-cool dwarfs in the Southern Galactic Plane, using data from the 2MASS and SuperCOSMOS Sky Surveys. Strong FeH and H$_2$O features in the near-infrared {\em JH}-band spectrum reveal characteristics seen in both mid-L, and L-T transition type dwarfs. We suggest these may be attributable to holes in the condensate cloud layers in the atmosphere of a single sub-stellar object, but cannot at present completely rule out the role of binarity as the possible cause. We also identify this object as a blue L dwarf, and explore the similar observable characteristics of these objects with those of the L-T transition. From this comparison we suggest that the temperature (and thus spectral type) at which the condensate cloud later begins to break-up/rain out, may be highly sensitive to small variations in metallicity. However, the {\em JH}-band spectrum of this object does not resemble that of the known L sub-dwarfs, and therefore extreme metal deficiency may not in fact be responsible for the discordant features. We estimate a spectral type of L9$\pm$1, and measure a large proper-motion of $\mu_{\mathrm{(tot)}}$=$1\farcs65\pm0\farcs03$ \mbox{yr$^{-1}$}. Also, a spectrophotometric distance of 8.2pc is estimated, possibly making this object the nearest easily observable single L-T transition object in the southern hemisphere.
\end{abstract}


\begin{figure*}
  \begin{center}
  \begin{minipage}{1.0\textwidth}
  \begin{center}
  \includegraphics[width=0.23\textwidth,angle=90]{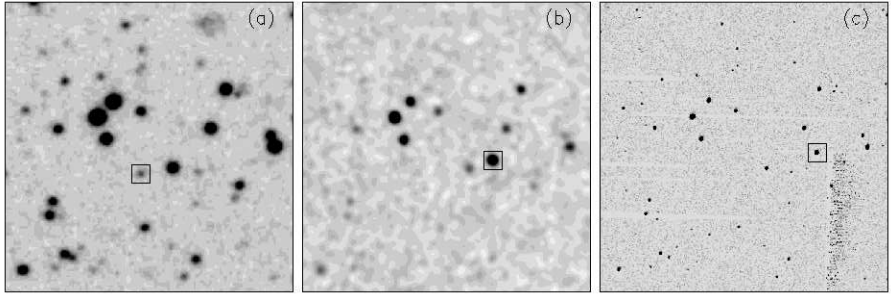}
    \caption{Discovery images for 2M1126-5003, showing from left to right: {\em (a)} SSS \Ivn~plate 1985 March 14$^{th}$, {\em (b)} 2MASS \Ks~image 1999 May 9$^{th}$, {\em (c)} SOFI {\em J}-band acquisition image 2006 April 8$^{th}$. All images are scaled to the same spatial resolution, 2\arcmin~on a side. North is up and East to the left. An SSS {\em R}-band image also exists but is not reproduced here as it is close in epoch to the \Ivn~image. The position of 2M1126 is identified in each image by a box, clearly showing the large proper-motion.}
    \label{f1}
  \end{center}
  \end{minipage}
  \end{center}
\end{figure*}

\section{Introduction}
Since the discovery of the first brown dwarfs Gliese 229B and Teide 1 in 1995 (\citealt{gl229b}; \citealt*{teide1}), the field of observational sub-stellar astrophysics has made significant advances in the classification and understanding of the nature of ultra-cool dwarfs (spectral types later than M6V), due mainly to the discovery of over 500 such objects from surveys such as the Two Micron All Sky Survey (2MASS; \citealt{Skrutskie}) and Sloan Digital Sky Survey (SDSS; \citealt{sdss}). These new ultra-cool dwarfs have observed properties that required an extension to the Harvard spectral classification system, to include two new spectral sequences; L and T  \citep[see][ for a detailed review]{kirk_review}.

The transition from M to L spectral types is marked by a disappearance of gas-phase TiO and VO, with a strengthening of metal hydrides (e.g., FeH, CrH, and CaH), H$_2$O bands, and alkali lines such as K{\sc\,i} and Na{\sc\,i}. As the temperature decreases through the L sub-types the \JK~colour reddens due to condensate formation in the photosphere.

The transition from L to T is characterised by the appearance of CH$_4$ absorption at $1.65$ and $2.2\mu m$, the depletion of condensate grains from the photosphere, along with deeper {\em JHK}-band H$_2$O absorption, and {\em HK}-band H$_2$ Collision Induced Absorption (CIA). Together these factors cause a reversal of the Near-InfraRed (NIR) colour trend resulting in these colours becoming bluer \citep[][, hereafter G02]{geballe}.

The observational properties of L-T transition objects (L9-T3, $\sim$1300K; \citealt*[][, hereafter BS06]{burrows}) have proved problematic to understand, due to complications introduced by the interplay of temperature, gravity, metallicity, and the physics of atmospheric dust clouds. These modify the NIR spectra, producing not only blue NIR colours, but also effects such as the {\em z}- to {\em J}-band brightening (\citealt[][, hereafter V04]{vrba}; \citealt{dahn}; \citealt*{tinney}; \citealt[][b, hereafter B06b]{burg_astroph06_1}), and the reappearance of FeH at $\sim$0.99um \citep[][ hereafter B02]{burg_lt}. As discussed by \citet[][, hereafter K04]{knapp} these effects appear to happen over a narrow range in effective temperature ({\em T}$_{eff}$), and may be caused by holes appearing in the condensate cloud layer, as suggested by B02. As such, single L-T transition objects should be rare, although unresolved binarity is known to be responsible for enhancing the numbers of early-T dwarfs from integrated light discoveries (B06b).

In this letter we report the discovery of a nearby L-T transition object, and discuss its observed spectral features in the context of: binarity, sub-solar metallicity/unusual gravity, and a single L-T transition object with observed properties akin to the previously discovered class of blue L dwarfs, identified by both \citet{knapp} and \citet[][, hereafter C06]{chiu}.

\section{Discovery of 2M1126}

\begin{table}
\begin{center}
\begin{minipage}{0.48\textwidth}
\begin{center}
 \caption{Observational parameters of 2M1126-5003}
 \label{t1}
 \begin{tabular}{lcc}
  \hline
  \hline
  Parameter & ~~~~~~~~~~~~~~ & Value \\
  \hline
  $\alpha _{J2000}$$^a$ & .............. & $11^{h} 26^{m} 39\fs91$ \\
  $\delta _{J2000}$$^a$ & .............. & $-50\degr 03\arcmin 55\farcs0$ \\
  2MASS~{\em J} & .............. & $14.00\pm0.03$ \\
  2MASS~{\em H} & .............. & $13.28\pm0.04$ \\
  2MASS~\Ks & .............. & $12.83\pm0.03$ \\
  {\em R$_C$$^b$} & .............. & $20.0\pm0.3$ \\
  {\em I$_C$$^b$} & .............. & $17.7\pm0.3$ \\
  $\mu_{\mathrm{(tot)}}${\em $^c$} & .............. & $1\farcs65\pm0\farcs03$ \mbox{yr$^{-1}$} \\
  $\mu_{\mathrm{(\alpha)}}${\em $^c$} & .............. & $-1\farcs58\pm0\farcs01$ \mbox{yr$^{-1}$} \\
  $\mu_{\mathrm{(\delta)}}${\em $^c$} & .............. & $0\farcs45\pm0\farcs01$ \mbox{yr$^{-1}$} \\
  Spectral~Type & .............. & L9$\pm$1 \\
  Distance{\em $^d$} & .............. & $8.2^{+2.1}_{-1.5}$ \mbox{pc}\\
  $V_{\mathrm{(tan)}}${\em $^e$} & .............. & $63^{+16}_{-12}$ \mbox{kms$^{-1}$} \\
  \hline
 \end{tabular}
\end{center}
 \medskip
 {\small {\em $^a$} 2MASS position (JD 2451308.49). {\em $^b$} Derived from the SuperCOSMOS \Rone~\&~\Ivn~plate magnitudes using the transforms from Bessell (1985). {\em $^c$} The uncertainty in the proper-motion is the percentage error derived from the residuals in transforming the SSS \Ivn~image, onto the 2MASS \Ks-band image geometry. {\em $^d$} Spectrophotometric distance using {\em M$_{K_s}$}$=13.27$ for an L9 derived from the absolute magnitude/spectral type relation of \citet[][c, astro-ph/0611505]{b06_2} converted to 2MASS photometry, with the uncertainty in distance reflecting the uncertainty in spectral subtype. {\em $^e$} Uncertainty in transverse velocity reflects the uncertainty in distance only.}
\end{minipage}
\end{center}
\end{table}

2MASS J11263991-5003550 (hereafter 2M1126) was identified during an ongoing programme to discover southern ultra-cool dwarfs (from late-M to the L-T transition) near to the Galactic Plane (GP; $|${\em b}$|$ $\lid$ 15\degr), the details of which will be discussed fully in a future paper. We note that 2M1126 was first catalogued in DENIS (DEep Near-Infrared Survey: \citealt{denis99}) as DENIS J112639.9-500355, observed at two epochs: 1999 April 6$^{th}$, and 1999 May 29$^{th}$. However, we chose to use the 2MASS catalogue identification from which this object was discovered.

Briefly, our programme utilises both the 2MASS and the SuperCOSMOS Science Archive \& Sky Survey (SSA and SSS respectively: \citealt*{hambly}), to define an optimal set of NIR and optical-NIR colour cut criteria, as well as Reduced Proper-Motion (RPM) criteria, where reliable optical counterparts can be found. Objects identified as being optically variable from the SSA dual epoch \Rone~and~\Rtwo~{\em R}-band data are also removed. Candidates are then visually inspected and crossed-matched with the on-line data bases of the NASA Extragalactic Database ({\sc NED}) and {\sc SIMBAD}. These procedures allow us to identify and remove contaminant objects including carbon giants, Long Period Variable stars (LPVs), T-Tauri stars, objects associated with nebulosity, galaxies, and optical plate defects$/$artefact's. These objects would otherwise overwhelm more simple photometric selections for ultra-cool dwarfs at such Galactic latitudes.

2M1126 stood out in our candidate selection sample due to its large proper-motion, evident between the optical (\Rone~and \Ivn)~plates and the 2MASS position. The proper-motion was measured using the 2MASS \Ks-band,~and SSS \Ivn-band images (baseline of 14.15 yrs) and was found to be $\mu_{\mathrm{(\alpha)}}$=$-1\farcs58\pm0\farcs01$ \mbox{yr$^{-1}$} and $\mu_{\mathrm{(\delta)}}$=$0\farcs45\pm0\farcs01$ \mbox{yr$^{-1}$}. The discovery \Ivn,~\Ks~images, and the most recent SOFI spectroscopic {\em J}-band acquisition image (see \S\ref{sobs}) are shown in Fig.~\ref{f1}, while the photometric and astrometric data are summarised in Table~\ref{t1}.

\section{Spectral Observation and Data Reduction}
 \label{sobs}
Follow-up NIR spectroscopy was obtained with the SOFI camera/spectrograph mounted on the ESO New Technology Telescope (NTT), on 2006 April 8$^{th}$ and 9$^{th}$. The blue low resolution {\em JH} grism was used with a 0\arcsec.6 slit width giving a resolving power of $\approx1000$, and a wavelength coverage of 0.95 to 1.65$\mu$m. The resulting dispersion was 6.96\AA pixel$^{-1}$. Standard calibration dome flats and xenon arcs where taken at the beginning and end of the night. An L5 template standard 2MASS J15074769-1627386 (hereafter 2M1507) was also observed with the same setup. We obtained a {\em J}-band acquisition image of 2M1126 while acquiring it on the slit. The targets were nodded along the slit to facilitate sky subtraction, with total integration times of 6x120secs for 2M1126, and 3x120secs for 2M1507. Observing conditions were good, with generally sub-arcsecond seeing (0\farcs5-1\arcsec), and the observations were carried out at an airmass $<$1.16. Telluric standard stars from Late-B to early-A spectral type were observed before and after the target acquisitions, and within 0.05-0.1 in airmass.

Image reduction and spectral extraction was achieved using standard packages within the {\sc iraf} environment, however, telluric correction and flux calibration was performed using the IDL based {\sc xtellcor\_general} routine provided by the NASA InfraRed Telescope Facility (IRTF), as part of the {\sc spextool} data reduction package (see \citet*{vacca} for details). This routine uses a high-resolution empirical model of Vega to construct a telluric correction spectrum with effective removal of the stellar absorption lines from telluric standard A0 stars, via interactive scaling of these features. Flux calibration was achieved within the same package using the known B and V magnitudes of the telluric standard (HIP044320), which were obtained from {\sc SIMBAD}, with a resulting F$(\lambda)$ flux scale.

As we compare our spectrum of 2M1126 to normalised composite spectra of known L and T dwarfs in this paper, we tested the robustness of our flux calibration by performing another telluric correction using a different standard taken several hours after the spectrum of 2M1126. We find no significant flux calibration errors across the whole spectrum, where the overall continuum level (avoiding the 1.11 and 1.4$\mu$m H$_2$O bands) varies by $<$3\% using the different telluric standards. However, we did find that using this temporally later telluric standard gave deeper H$_2$O band absorption at 1.11 and 1.4$\mu$m in our spectrum of 2M1126. This had the effect of increasing the derived spectral type by $+$1 when using spectral indices defined for these H$_2$O bands (see \S\ref{nis} below). However, note that this level of uncertainty is equal to our final estimated spectral type uncertainty for 2M1126 (see Table~\ref{t2}).

\section{Distance and Kinematics of 2M1126}

\begin{figure}
\begin{center}
\includegraphics[width=0.48\textwidth,angle=0]{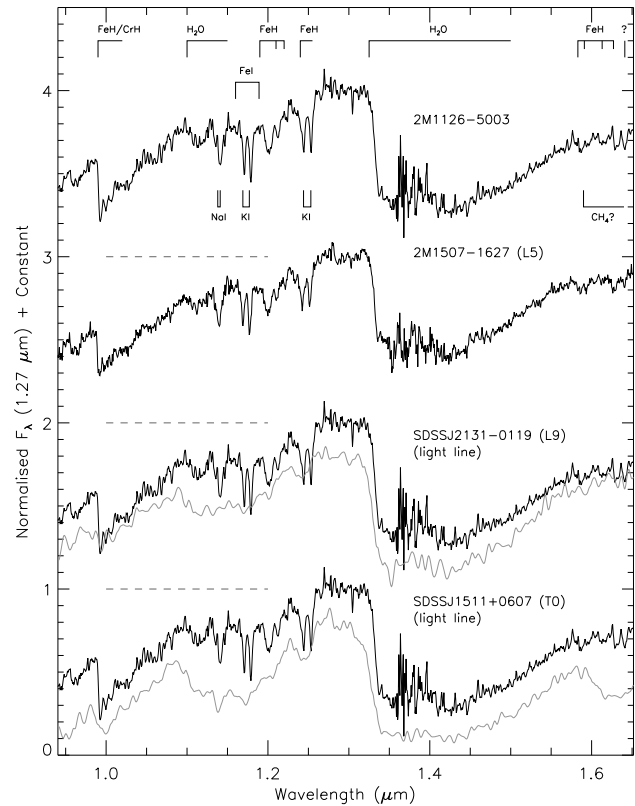}
 \caption{The spectrum of 2M1126-5003 (top) shown with spectral feature identifiers. Also shown are the L5 (2M1507) from this work 2$^{nd}$ from top, and two comparison L9 and T0 spectra (SDSS J2131-0119 3$^{rd}$ from top, and SDSS J1511+0607 bottom plot; both from C06), plotted with light grey lines and offset from the spectra of 2M1126 (solid lines) by -0.2 in normalised units for clarity. The zero level for each 2M1126 spectrum is indicated by a dashed line.}
 \label{f2}
\end{center}
\end{figure}

We derive a spectrophotometric distance of $8.2^{+2.1}_{-1.5}$pc based on the \Ks(MKO)~absolute magnitude$/$spectral type relation of \citet[][c, astro-ph/0611505]{b06_2}, which contain decomposed binary data spanning the L-T transition. We converted this relation onto the 2MASS photometric system, and estimated an absolute magnitude of M{\em $_{K_s}$}$=13.27$ for 2M1126, based on the spectral type of L9$\pm$1 that we estimate in \S\ref{slt}. The uncertainty in the distance reflects the uncertainty in the spectral sub-type, as well as the RMS error in the M{\em $_{K_s}$}$/$spectral type relation. We have assessed the parallax distances (where available), and spectrophotometric distances (as estimated for 2M1126) of all currently known L-T transition objects (L9 to T3; {\sc dwarfarchives.org}). If the distance of $8.2$pc is confirmed by parallax, then 2M1126 is the fourth nearest L-T transition object, with only $\bepsilon$ Indi Ba (T1), DENISPJ0255-4700 (L9 at $\sim$4.5pc; \citealt{martin}, \citealt{denis255}, and \citealt{burg_uni_class}), and IBIS J013656+093347 (T2.5: \citealt{2nd_north_Td}) being closer. Given that $\epsilon$ Indi Ba is a member of a close binary (resolved with both AO and HST imaging), and IBIS0136+0933 is a northern object, then 2M1126 is potentially the nearest easily observable single L-T transition object in the southern hemisphere. It is not known whether DENISPJ0255-4700 is a binary. However, 2M1126 is the nearest blue L dwarf currently identified.

The corresponding tangential velocity of $V_{tan}=63$kms$^{-1}$ for 2M1126 is at the upper end of the range typical for late-L dwarfs (V04), and possibly suggestive of membership of the intermediate Galactic disk population, implying that 2M1126 may have a slightly lower than solar [Fe/H].

\begin{table*}
\begin{center}
\begin{minipage}{1.0\textwidth}
\begin{center}
 \caption{Spectral Indices for 2M1126.}
 \label{t2}
 \begin{tabular}{lccccccccc}
  \hline
  \hline
  Assigned & H$_2$O$^a$ & H$_2$OB$^b$ & H$_2$OA$^c$ & H$_2$OB$^c$ & H$_2$O-{\em J}$^d$ & H$_2$O-{\em H}$^d$ & {\em z}-FeH$^c$ & {\em J}-FeH$^c$ & FeH-b$^e$ \\
  Type & $1.5\mu m$ & $1.5\mu m$ & $1.343\mu m$ & $1.456\mu m$ & $1.15\mu m$ & $1.40\mu m$ & $0.988\mu m$ & $1.20\mu m$ & $0.99\mu m$ \\
  \hline
  L9$\pm 1^*$ & 1.77(L8) & 0.56(L6.5) & 0.36(L8.5) & 0.53(L7.5) & 0.72($<$L8) & 0.65(L9.5) & 0.54(L0-L5) & 0.87(L0-L5) & 2.11($\sim$L5) \\
  \hline
 \end{tabular}
\end{center}
 \medskip
 {\small
 {\em $^*$} Final spectral type assigned from both the spectral template fitting and the H$_2$O spectral indices (see \S\ref{slt}).\\
 Spectral classification schemes from: {\em $^a$} \citet{geballe}; {\em $^b$} \citet{reid} H$_2$O$^B$ index; {\em $^c$} \citet{mcLean} H$_2$OA, H$_2$OB \& FeH indices; {\em $^d$} \citet{burg_uni_class} unified NIR classification; {\em $^e$} \citet{kirkp} FeH-b $\sim$0.99$\mu$m band index used by \citet{burg_lt} (see Fig.\ref{f6}). Implied spectral types given in brackets are rounded to the nearest half sub-type.}
\end{minipage}
\end{center}
\end{table*}

\begin{table*}
\begin{center}
\begin{minipage}{1.0\textwidth}
\begin{center}
 \caption{Spectral Indices for the L5 comparison standard 2M1507.}
 \label{t3}
 \begin{tabular}{lcccccccc}
  \hline
  \hline
  Assigned & H$_2$O$^a$ & H$_2$O$^a$ & H$_2$OB$^b$ & H$_2$OA$^c$ & H$_2$OB$^c$ & z-FeH$^c$ & J-FeH$^c$ & FeH-b$^d$ \\
  Type & $1.2\mu m$ & $1.5\mu m$ & $1.5\mu m$ & $1.343\mu m$ & $1.456\mu m$ & $0.988\mu m$ & $1.20\mu m$ & $0.99\mu m$ \\
  \hline
  L5$^*$ & $<$T0 & 1.58(L5) & 0.64(L4.5) & 0.49(L5) & 0.63(L5) & 0.56(L0-L5) & 0.85(L0-L5) & 1.79($\sim$L5) \\
  \hline
 \end{tabular}
\end{center}
 \medskip
 {\small
 {\em $^*$} Final spectral type assigned from the H$_2$O and FeH spectral indices (see \S\ref{slt}).\\
 Spectral classification schemes from: {\em $^a$} \citet{geballe}; {\em $^b$} \citet{reid} H$_2$O$^B$ index; {\em $^c$} \citet{mcLean} H$_2$OA, H$_2$OB \& FeH indices; {\em $^d$} \citet{kirkp}, FeH-b $\sim$0.99$\mu$m band index used by \citet{burg_lt} (see Fig.\ref{f6}). Implied spectral types given in brackets are rounded to the nearest half sub-type.}
\end{minipage}
\end{center}
\end{table*}

\section{Analysis of 2M1126}
\subsection{Near-Infrared Spectrum}
 \label{nis}

The spectrum of 2M1126 is shown in Fig.~\ref{f2} (top plot), as well as an L5 (2M1507 observed by us; 2$^{nd}$ from top). Also shown in this figure are two comparison L9 and T0 spectra (SDSS J2131-0119 3$^{rd}$ from top, and SDSS J1511+0607 bottom plot; both from C06), plotted with lighter weight lines offset from the 2M1126 spectra (solid lines) by -0.2 in normalised units.

Due to the presence of the deep H$_2$O band at 1.4$\mu$m along with H$_2$O absorption at 1.11$\mu$m, it is immediately obvious that 2M1126 is a late-type ultra-cool dwarf. Other prominent features in the spectra are the strong hydride bands of FeH$/$CrH at $\sim$0.99$\mu$m \citep{wf}, FeH in the {\em J}-band at 1.19, 1.22 \& 1.24$\mu$m (\citealt{jones}; \citealt{reid}; \citealt{mcLean2000}), and in the {\em H}-band at 1.583, 1.591, 1.613 and 1.625$\mu$m (\citealt{wh}: \citealt{crdv}: \citealt{reid}). Also indicative of late-type are the strength of the two K{\sc\,i} doublets ($\sim$1.175 and $\sim$1.25$\mu$m), and the Na{\sc\,i} lines at $\sim$1.14$\mu$m. Fe{\sc\,i} can be seen at $\sim$1.160 and 1.189$\mu$m \citep*[][ : hereafter C05]{Cushing}. There is also a prominent absorption feature located at 1.641$\mu$m which has been previously seen in VB10. This unidentified feature does not appear to be caused by FeH, and is enhanced in 2M1126 compared with VB10 \citep{crdv}. The FeH absorption in 2M1126 is particularly strong especially at 0.99$\mu$m, and is also unusually prominent at around 1.6$\mu$m for such a late-type object.

For 2M1126, H$_2$O spectral indices were calculated under both the unified scheme of \citet[][a]{burg_uni_class} and the schemes of G02, M03, and \citet{reid}, while the FeH indices are from M03 and \citet{kirkp}. These same schemes were also used for 2M1507 except those of \citet[][a]{burg_uni_class}. The index values and implied spectral types are summarised in Table~\ref{t2} for 2M1126, and Table~\ref{t3} for 2M1507. The L5 spectral type we find for the previously known 2M1507 is in agreement with the published value (M03). For 2M1126, the average of all the H$_2$O indices gives an estimated spectral type of L8, while the average of the two \citet[][a]{burg_uni_class} H$_2$O indices implies a spectral type of L9. The {\em z}-FeH, FeH-b, and {\em J}-FeH indices show surprisingly strong FeH absorption, and comparing the FeH strength with 2M1507 indicates a spectral type in the $\sim$L3 to L5 range. Given these discrepant spectral types obtained from different features, it is thus not possible to assign a spectral type to 2M1126 based on all of these spectral indices alone.

This apparent `duality' in the nature of the spectrum of 2M1126 is also evident from the general spectral morphology. It may be seen in Fig.~\ref{f2} that the spectral region from 1.19 to 1.30$\mu$m shows similar FeH band absorption strengths, and K{\sc\,i} line widths, to those seen in early- to mid-L dwarfs. The 0.99$\mu$m FeH$/$CrH band also appears to be even stronger than for 2M1507. However, it may also be seen that both the comparison L9 and T0 spectra in Fig.~\ref{f2} are comparable to 2M1126 in the wavelength range $\sim$1.3 to 1.6$\mu$m, which contains the 1.4$\mu$m H$_2$O absorption.

Possible scenarios which may highlight the nature of 2M1126 are now discussed.

\subsection{An Unresolved L-T Binary?}
 \label{LTbin}

The PSF of 2M1126, which has a FWHM of $0\farcs59\pm0\farcs01$ measured from the SOFI acquisition image, shows no hint of binarity when compared with eleven neighbouring objects, which have a mean FWHM of $0\farcs57\pm0\farcs02$. At a distance of 8.2pc and using the difference between the minimum and maximum 3$\sigma$ uncertainty applied to the FWHM measurements of 2M1126 and neighbouring objects, we find a lower limit to the detectability of any on unresolved binarity of $\ga 0\farcs11$. This implies a lower limit to the physical separation of any binary components which we should be able to detect in  this image of $\ga 0.9$AU (assuming a favourable orbital phase). This is below the currently determined (possibly resolution limited) peak of the T dwarf binary separation distribution of $\sim$4AU (B06b). However, it is possible that 2M1126 may still be a more closely separated binary system.

\begin{figure}
\begin{center}
\includegraphics[width=0.48\textwidth,angle=0]{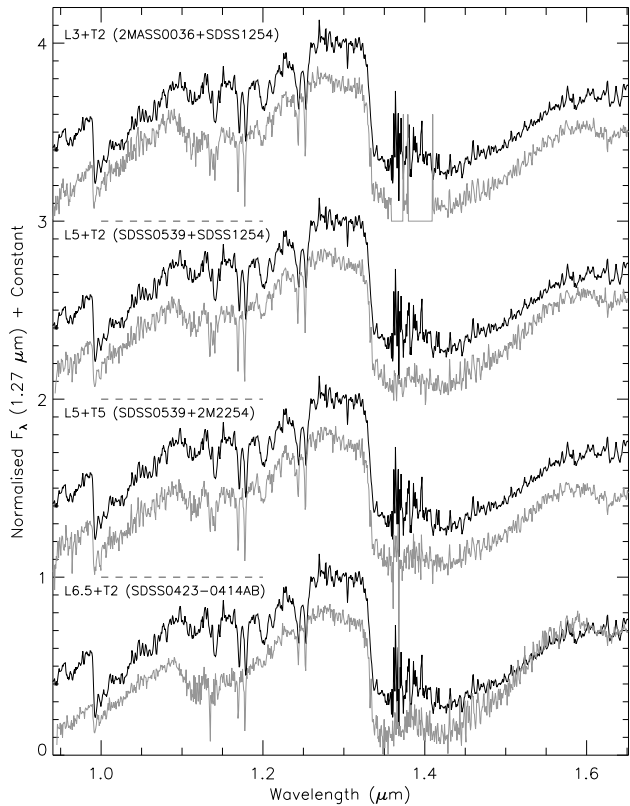}
 \caption{The spectrum of 2M1126-5003 (solid dark lines), with all over-plotted spectra shown as light grey, and offset below 2M1126 in each case by -0.2 in normalised units for clarity. The top three plots represent our best attempts to match composite L \& T spectra to that of 2M1126. For the L3+T2 and L5+T2, the L \& T components where first normalised (at 1.27$\mu m$), and then summed without any scaling. For the L5+T5, both components were also normalised as above and then summed, but after the T5 component was scaled further by a factor of 0.4. The bottom plot is of the known resolved binary SDSS J0423-0414 (see text in \S\ref{LTbin} for details). The L3 and L5 spectra are from C05, while the T2 and T5 spectra are from M03. The zero level for each 2M1126 spectrum is indicated by a dashed line.}
 \label{f3}
\end{center}
\end{figure}

We find that composite L and T spectra of known dwarfs do not accurately reproduce the flux above and below $\sim$1.25$\mu$m at the same time, due primarily to: an enhanced {\em zJ}-band flux, the FeH band strength at 0.99$\mu$m, and lack of 1.65$\mu$m CH$_4$ absorption in 2M1126. The FeH absorption and {\em J}-band flux can be made to fit better with an appropriate scaling of the `T' component (i.e., reducing the contribution of the `T' component), but at the expense of a good fit to the 1.4$\mu$m H$_2$O absorption. In Fig.\ref{f3} we show three combinations which we found gave the closest match to 2M1126 comprising an L3+T2, L5+T2, and an L5+T5 pair (see caption in Fig.\ref{f3} for details of the scaling used). Given our confidence in the overall continuum shape of our 2M1126 spectrum (see \S\ref{sobs}), the differences seen between our best fitting combinations, and 2M1126, appear significant.

We also compare the spectrum of 2M1126 with that of the integrated light spectrum of the known binary SDSS J0423-0412, which was originally spectral typed in the NIR (from an integrated spectrum) as T0 by G02, and was found to have {\em z}-FeH and {\em J}-FeH indices consistent with this type (M03), unlike for 2M1126. \citet{cruz} originally suggested that SDSS0423 was a single L-T transition object, however, it was subsequently resolved as a binary with components of spectral type L6.5 and T2 by B06b. As these components of SDSS0423 are of similar spectral type to one of our closest fitting (but unsuccessful) composite spectra (the L5+T2), it then appears less likely that 2M1126 is an unresolved binary with component spectral types of mid-L and early-T. This can be seen by the poor fit of the integrated light spectra of SDSS0423 when compared with the spectra of 2M1126 (see Fig.\ref{f3}, especially the FeH features, H$_2$O at 1.11$\mu$m, CH$_4$ at 1.6$\mu$m, and the {\em zJ}-band flux levels. 

As stated in \citet{cruz}, a late-L$/$mid-T binary system would still be expected to have a relatively red composite \JK~colour (although bluer than the colour of the mid-L primary alone), due to the L dwarf dominating the {\em K}-band flux as for SDSS0423 (\JK~$=1.51$; M03). Although 2M1126 is relatively red it is significantly bluer than SDSS0423, further suggesting that 2M1126 is not a binary system. However, we cannot completely rule out the possibility that 2M1126 is an unresolved binary.

\subsection{Sub-solar Metallicity?}
 \label{subsol}

Strongly enhanced collision-induced H$_2$ absorption which depresses the {\em HK}-band flux (see \citealt{saumon94}; \citealt*{borysow97} for details), strong metal hydride absorption, and the blue \JK~colour seen in L-type sub-dwarf spectra are indicative of extreme metal deficiency, as has been shown for 2MASS J05325346+8246465, and 2MASS J16262034+3925190 (\citealt{burg_subd}; \citealt{2nd_burg_subd} respectively). There is no significant sign of H$_2$ induced CIA suppressing the {\em H}-band spectrum of 2M1126, compared with that of the L9 and T0 shown in Fig.~\ref{f2}. However, the \JK~colour of 1.17 for 2M1126 is significantly bluer than normally seen for late-L type dwarfs (assuming 2M1126 is an L9 see \S\ref{slt} below).

\begin{figure}
\begin{center}
\includegraphics[width=0.48\textwidth,angle=0]{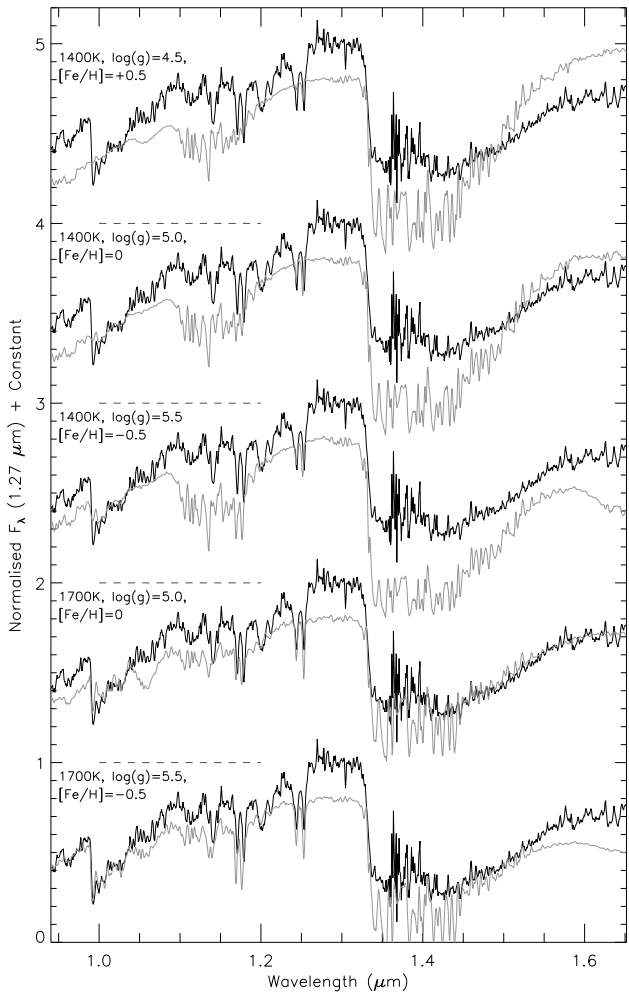}
 \caption{Model spectra from BS06 over-plotted in light grey on 2M1126 (solid lines), and offset from the spectra of 2M1126 by -0.2 in normalised units in each case. These show how the FeH and H$_2$O absorption and the {\em zJH}-band flux levels, might be expected to change as a function of temperature, metallicity and gravity. Top three model spectra are for an {\em T}$_{eff}$ of 1400K, $\log(g)=\{4.5,5.0,5.5\}$, and [Fe$/$H]$=\{+0.5,0,-0.5\}$. Bottom two model spectra are for an {\em T}$_{eff}$ of 1700K, $\log(g)=\{5.0,5.5\}$, and [Fe$/$H]$=\{0,-0.5\}$. All model spectra use a forsterite particle size of $100\mu$m. See \S\ref{subsol} for details. The zero level for each 2M1126 spectrum is indicated by a dashed line.}
 \label{f4}
\end{center}
\end{figure}

\begin{figure}
\begin{center}
\includegraphics[width=0.48\textwidth,angle=0]{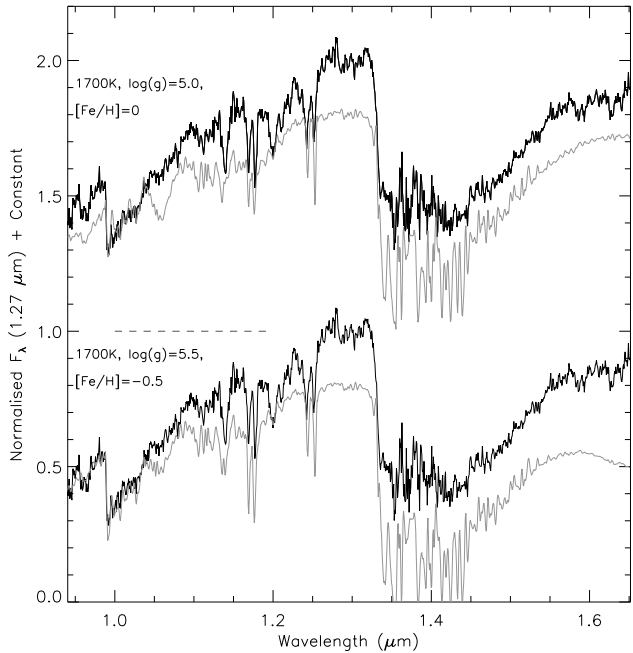}
 \caption{Model spectra from BS06 over-plotted in light grey on L5 2M1507 (solid line in both plots), and offset by -0.2 in normalised units in each case. The model spectra are for an {\em T}$_{eff}$ of 1700K, $\log(g)=\{5.0,5.5\}$, and [Fe$/$H]$=\{0,-0.5\}$. The model spectra use a forsterite particle size of $100\mu$m. See \S\ref{subsol} for details. The zero level for the top 2M1507 spectrum is indicated by a dashed line.}
 \label{f5}
\end{center}
\end{figure}

At present models of ultra-cool dwarf atmospheres do not accurately represent the observable parameters of objects inhabiting the L-T transition (L8-T3). Although this precludes a direct meaningful fit of parameters for such models with 2M1126, we used the latest Burrows models (BS06) to internally compare these models against 2M1126, and assess the theoretical effects of metallicity and gravity. We do this for the case of two different effective temperatures to see if the deep FeH absorption features we see in 2M1126 can be adequately reproduced. The models of BS06 do not include partial cloud coverage across the L-T transition, but they do contain updated 0.99$\mu$m FeH/CrH gas-phase molecular opacities. In Fig.~\ref{f4} we over-plot a total of five model spectra with 2M1126 for the case of two temperatures: (i) {\em T}$_{eff}=$1400K: low gravity and high metallicity, `normal' gravity and solar metallicity, and high gravity with low metallicity; and (ii) two model spectra for an {\em T}$_{eff}=$1700K: `normal' gravity and solar metallicity, and high gravity with low metallicity. In Fig.~\ref{f5} we also show model spectra over-plotted on the L5 2M1507 for `normal' gravity and solar metallicity, and high gravity with low metallicity, both for an {\em T}$_{eff}= $1700K.

A comparison of the model spectra with 2M1126 suggests that the higher temperature of {\em T}$_{eff}= $1700K with a metallicity of [Fe$/$H]$=-0.5$, and $\log(g)=5.5$ gives the best match to the FeH absorption, and the K{\sc\,i} and Na{\sc\,i} alkali absorption lines in the {\em zJ}-band. However, the overall relative flux levels across the spectrum of 2M1126 appears to be reproduced better by the {\em T}$_{eff}=$1400K model, again with a metallicity of [Fe$/$H]$=-0.5$, and $\log(g)=5.5$. For 2M1507, we see in Fig.~\ref{f5} that the $\log (g)=5.0$, and solar metallicity gives the best overall fit, however, a lower metallicity ([Fe$/$H]$=-0.5$) is needed to explain the FeH in 2M1507 absorption using this model.

Given that there is no apparent sign of significantly enhanced H$_2$ induced CIA in our {\em H}-band spectrum, and how dissimilar the spectrum and \JK~colour of 2M1126 are, compared with the those of the two L sub-dwarfs mentioned above, then if 2M1126 has a low metallicity it is unlikely to be significantly so, and is therefore not an L sub-dwarf. In this case the discordant {\em z}- and {\em J}-band features seen in 2M1126, and especially the strength of the FeH indices which give values consistent with early- to mid-L types, may possibly be linked to physical processes occurring in the atmosphere. In \S\ref{slt} below we will discuss what atmospheric processes could be responsible for the discordant features seen in 2M1126, and in particular the cloud clearing model of B02. We note that the proposed rapid rain out model proposed by K04 in which the sedimentation efficiency begins to increase once the atmosphere has cooled to some critical temperature ({\em T}$_{eff} \simeq$ 1300K) may possibly produce similar effects, as the cloud deck becomes thinner. However, due to the strength of the FeH absorption seen in 2M1126 we choose to concentrate on the cloud clearing/break-up mechanism, which may give a better explanation for the resurgence of FeH over the L-T transition.

\subsection{A Single L-T Transition Object?}
 \label{slt}

The reasonable fit of the template L9 and T0 in Fig.\ref{f2} to 2M1126 over the 1.4$\mu$m H$_2$O band spectral region (as well as the excellent fit of the L9.5 blue L dwarf SDSS0805+4812 seen in Fig.\ref{f7}), suggests an effective temperature of $\sim$1300-1400K (V04) for 2M1126, despite the much earlier spectral type suggested by the {\em z}-FeH and {\em J}-FeH indices. As a single object we therefore estimate a spectral type of L9 for 2M1126 based on this template fitting and the average of the \citet[][a]{burg_uni_class} 1.14 and 1.40$\mu$m H$_2$O band spectral indices, with an uncertainty of $\pm$1 sub-type to reflect the uncertainty in all the H$_2$O band spectral indices.

As a single object, the strong presence of FeH in 2M1126 is consistent with the scenario suggested by B02, in which holes are postulated to appear in the condensate cloud layer allowing {\em zJ}-band flux to escape from deeper, higher temperature regions, where the FeH abundance would be expected to be larger (BS06). As was shown in \S\ref{subsol} this scenario is also alluded to by the comparison of the best fitting model of BS06, to the {\em zJ}-band FeH absorption features seen in 2M1126 which appears to require a temperature of {\em T}$_{eff}$ of 1700K, and lower metallicity to explain the FeH absorption strength.

\begin{figure}
\begin{center}
\includegraphics[width=0.5\textwidth,angle=0]{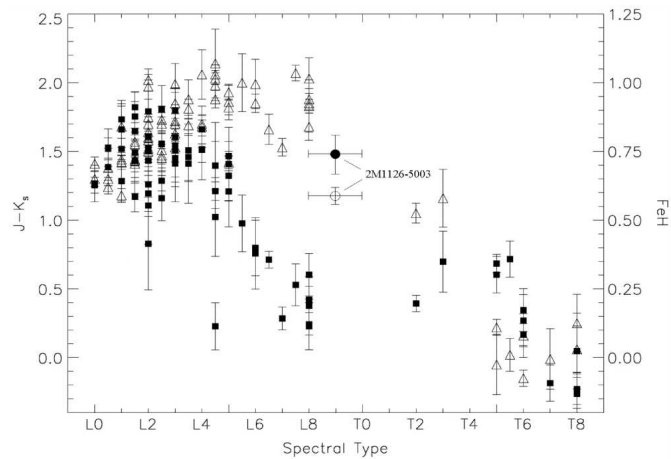}
 \caption{Reproduced from B02 (fig.3; filled squares are the $\ln (FeH$-$b)$ index, and open triangles 2MASS \JK~colour), with FeH and \JK~data for 2M1126 (filled, and open circles respectively) added assuming a spectral type of L9$\pm$1. The strength of the enhanced 0.99$\mu$m FeH absorption compared to other late-L and early-T dwarfs is evident, while the \JK~value fits in well with the trend.}
 \label{f6}
\end{center}
\end{figure}

Evidence for this may be found by comparing the \JK~colour and FeH absorption versus spectral type for 2M1126, with other L and T dwarfs as shown in Fig.\ref{f6}, which is reproduced from fig.3 in B02 with our data for 2M1126 added (also see \citealt{kirkp}). The 0.99$\mu$m FeH strength weakens rapidly from mid- to late-L, where \JK~becomes redder. For early- to mid-T dwarfs (as \JK~becomes bluer) FeH is seen to persist at a low level, with some objects showing a slight resurgence in FeH absorption compared to the latest L dwarfs. If holes are appearing in the atmospheric dust cloud layers for early-T dwarfs, then this could explain the resurgence of FeH absorption for such objects. But, as may be seen in Fig.\ref{f6}, the strength of the FeH absorption ($\ln (FeH$-$b)$ in Table.\ref{t2}) for 2M1126 is unusually strong compared to the other L-T transition objects. Interestingly, both the T2 and T3 shown in Fig.\ref{f6} that show much weaker FeH absorption than 2M1126 (SDSS1254-0122 and SDSS1021-0304 respectively), have H$_2$O and FeH spectral indices which show no scatter in their derived spectral types (M03), and the T3 has also been resolved as a binary by B06b.

So what could be causing the FeH absorption to be so much stronger in 2M1126 than in the other late-L and early-T dwarfs which show a reappearance of FeH? A look at fig.9 in BS06 which plots temperature-pressure profiles for {\em T}$_{eff}$ ranging between 2200K and 700K, shows the relative positions of the radiative and convective zones to that of the cloud bases and photosphere, for their ultra-cool dwarf model atmospheres. For early- to mid-L dwarfs the strong FeH arises from the hotter radiative photosphere, by late-L the photosphere now lies in the outer convection zone among the upper cloud deck, which is being depleted of FeH due to condensate formation. At the beginning of the L-T transition the convection zone and upper cloud deck recedes rapidly from the photosphere, over a small range in {\em T}$_{eff}$. The photosphere now begins to reside in an outer radiative zone which becomes progressively dominant over the emergent spectrum with decreasing temperature, as its optical depth increases through the mid- to late-T dwarfs, and the \JK~colour swings to the blue. The point at which the cloud deck begins to break-up across the L-T transition should thus dictate the strength of the resurgence of the FeH absorption features, and the level of the flux enhancement in the {\em z}- and {\em J}-bands. The further the cloud deck$/$convection zone recedes below the increasingly cooler radiative dominated photosphere, the less contribution there will be from these lower layers.

Thus, given the B02 cloud hole interpretation, such a strong enhancement of FeH seen in the spectra of 2M1126 could possibly be due to an early cloud break-up relative to the L-T transition spectral type. This would allow even a small hole filling factor to contribute significantly to the spectrum of 2M1126, due to flux from deeper, much warmer, layers than would normally be seen for early-T dwarfs, in which the reappearance of FeH absorption is far less prominent.

As the \JK~colour for 2M1126 is not yet very blue (1.17), a substantial upper radiative zone (with a large optical depth) has presumably not yet developed above the convection zone/cloud tops. Thus, it seems plausible that an early cloud break-up is the possible cause of the strong FeH absorption, and the flux enhancement of the {\em z}- and {\em J}-bands (compared with the over-plotted template L9 and T0 in Fig.\ref{f2}). With such an unambiguous and unusual strength of FeH, and no significant sign of H$_2$ CIA suppression in the {\em H}-band, cloud disruption$/$break-up could therefore play a significant role in the atmospheric physics responsible for the features seen in 2M1126, and possibly of other L-T transition objects.

An early cloud break-up could be a normal occurrence for L-T transition objects (i.e., at $\sim$T0 for $\sim$solar metallicity), but due to the small {\em T}$_{eff}$ range over which this transition appears to occur, and subsequent rapid evolution through this phase (as implied by the poor fit of the models of BS06 across this transition), genuinely single L-T transition objects should therefore be rare. This may explain why no other object has been identified as an L-T transition object which shows such pronounced FeH enhancement as 2M1126, especially in light of recent work showing that samples of early-type T dwarfs could be significantly contaminated by L$/$T crypto-binarity (BS06). However, there is one known object SDSS0805+4812 (an L9.5$\pm$1.5: K04 and C06) which does show strong FeH absorption, but was identified as a blue L dwarf by K04. This object (which will be discussed in \S\ref{blueld} below) is very similar to 2M1126 both spectroscopically, and in terms of its NIR colours.

As suggested by \citet{crdv} the reappearance of FeH in the {\em J}- and {\em H}-bands in early-T dwarfs is not normally seen, due to strong CH$_4$ absorption masking these spectral regions. However, in 2M1126 the apparent absence of CH$_4$ allows the reappearance of FeH to be easily identified in these bands, which also display a strength similar to, or greater than that seen in the L5 2M1507 like that seen for the 0.99$\mu$m FeH absorption.

Although there is no sign of CH$_4$ at 1.65$\mu$m in 2M1126, C06 identified an L7.5$\pm$2.5 dwarf (SDSS1025+3212) with scattered spectral indices in which CH$_4$ at 1.65$\mu$m is variable. This object also has spectral features usually associated with both T0 and mid-L dwarfs. Therefore CH$_4$ at 1.65$\mu$m may also be variable in 2M1126, and possibly intrinsic to the atmospheres of objects undergoing the initial period of the L-T transition.

\section{Blue L dwarfs and the L-T transition}
\label{blueld}

\begin{figure}
\begin{center}
\includegraphics[width=0.48\textwidth,angle=0]{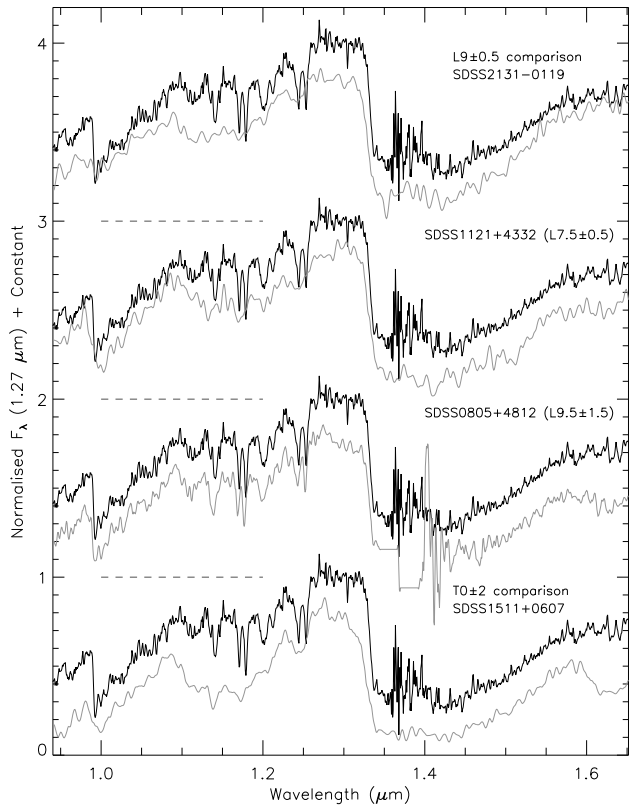}
 \caption{The spectra of two blue L dwarfs shown 2$^{nd}$ and 3$^{rd}$ from top (SDSS1121+4332 and SDSS0805+4812 respectively), bracketed top and bottom by typical L9 and T0 spectra (SDSS2131-0119 and SDSS1511+0607 respectively). Over-plotted spectra shown as light grey and plotted -0.2 lower in normalised F$_\lambda$ units relative to 2M1126 (solid dark lines) in each case. All spectra and spectral classifications taken from C06. The zero level for each 2M1126 spectrum is indicated by a dashed line.}
 \label{f7}
\end{center}
\end{figure}

A number of L dwarfs with unusually blue \JK~colours have been previously identified, and range in spectral type from L1 to L9.5, typically with an uncertainty of more than one sub-type as derived from their spectral indices. K04 identified four, which they show to have strong FeH and K{\sc\,i} features similar to early-L dwarfs, while their H$_2$O band strengths are more like those of late-L types. They also find their spectral indices show a large scatter in implied spectral type, with the {\em H}-band indices implying a late type of L9 to T1, while the {\em K}-band implies $\sim$mid-L. They suggest that this supports the idea that in these objects we are seeing to greater depths free from condensates. C06 also identified three such blue L dwarfs with the spectra of two of these shown in Fig.~\ref{f7}. Another two blue L dwarfs (2MASS J1300425+191235 and 2MASS J172139+334415) were discovered by \citet{gizis00} and \citet{cruz03} respectively, both having \JK~colours of $\sim$1.1. As can be seen in Fig.~\ref{f7} the two blue L dwarf spectra show very similar overall morphology to 2M1126, particularly that of SDSS J0805+4812 (an L9.5, C06: hereafter SDSS0805) which has very similar absorption features and NIR colours. C06 suggest that the strong H$_2$O and FeH absorption bands in these three objects may be due to sub-solar metallicity, and$/$or thinner condensate cloud layers. More generally 2M1126 shares many similarities with the blue L dwarfs such as: similar H$_2$O band depths at 1.11 and 1.4$\mu$m, strong {\em zJ}-band FeH absorption especially at 0.99$\mu$m, enhanced {\em zJ}-band flux, and the relative flux level of the sharp 1.1$\mu$m {\em J}-band peak to that of the {\em H}-band peak (typical for objects across the L-T transition).

Another similarity between SDSS0805 and 2M1126 is that SDSS0805 has an optical spectral type of L4 \citep{hawley} while 2M1126 has optical spectral type of L4.5 (Burgasser 2006, private communication), both of which are significantly earlier than their NIR derived types. Thus, these similarities suggest that 2M1126 and SDSS0805 are both L-T transition objects as well as being blue L dwarfs. 

In the single object interpretation the apparent similarity of SDSS0805 with 2M1126, as well as the other blue L dwarfs of earlier spectral type, may result from these objects all undergoing a transition to clearer atmospheres. Thus, the blue L dwarfs possibly represent objects with a similar transitional phase taking place in their atmospheric physics/chemistry, to that seen at the usual L-T transition. This could result in the bluer \JK~colours seen for both the blue L dwarfs, and those of the L-T transition (i.e., {\em T}$_{eff}$ of $\sim$1400K).

BS06 show from their models that while the effect of gravity on the M{\em $_J/$}\JK~space across the L-T transition is weak, the metallicity dependence is strong. They also show that sub-solar metallicity L dwarfs are expected to swing to the blue in \JK~at an earlier spectral type than $\sim$solar metallicity ones. It would appear that the spectral type (and temperature) at which the condensate cloud layer begins to break-up$/$rain out for ultra-cool dwarfs (and thus the transition to clearer dust free atmospheres), may be highly sensitive to metallicity. Therefore, a small variation in [Fe$/$H] may shift these transitional atmospheric processes to earlier spectral types, possibly explaining the blue L dwarfs (i.e., having a slightly lower metallicity than seen for objects at the usual L$/$T boundary). It would seem that given their observed properties the blue L dwarfs fit in naturally with this picture, if one accepts them as transitional objects such as those undergoing the L-T transition.

The \JH~$/$\HK~two colour diagrams plotted in figures 4 and 6 of K04 show the colours of the four blue L dwarfs identified by these authors, relative to other representative L and T dwarfs. These blue L dwarfs lie just at the beginning, and slightly below (bluer in \JH~) the T sequence, well away from the mid- to late-L dwarf sequence. The slightly bluer \JH~colours may possibly be explained by an enhanced {\em J}-band flux from much warmer FeH rich lower layers, before the atmosphere begins to cool appreciably through the T sequence. Interestingly, 2M1126 if plotted on the same fig.6 in K04 (with colours of \JH~= 0.71, and \HK~= 0.46) lies at the position consistent with an $\sim$L9-T0 spectral type.

If one takes the NIR H$_2$O band strengths, and also the fact that there is no appreciable sign of H$_2$ CIA suppression in the {\em H}- and {\em K}-bands, as a indicator of temperature and spectral type for these objects (including 2M1126), then it would seem reasonable to suggest that the blue L dwarfs (and 2M1126) are also transitional objects, such as those at the L$/$T boundary. Therefore, the blue L dwarfs could potentially form part of the natural sequence between the L-type dust dominated atmospheres, and the optically thin T dwarf ones, as seen in the NIR two-colour diagram mentioned above, but occurring at an earlier spectral type dependant on metallicity.

The cloud clearing model of B02 suggests that the condensate clouds clear along almost isothermal tracks (see fig.1 in B02). However, comparisons with empirical data imply that although the L-T transition (defined by a swing to bluer \JK~colour) is initiated at an {\em T}$_{eff}$ of $\sim$1200K, there is some intrinsic scatter in this around temperature. This scatter is stated as being likely due to the influence of gravity, rotation, and metallicity on the cloud coverage by the authors of B02, which shows some mid- to late-L dwarfs occupying the region of partial cloud coverage ($\sim$60 to 80\%) in fig.1 of B02, with a {\em T}$_{eff}$ of 1400 to 1600K indicated by their model tracks.

If our suggestion is correct, then the spread in the spectral indices seen for the blue L dwarfs (and 2M1126), along with their discordant spectral features and morphology as we have discussed in \S\ref{slt}, may therefore reflect the transition to clearer atmospheres by the condensate cloud break-up, in the same way as for objects undergoing the L-T transition.

\section{Summary and Future Work}

2M1126 is a nearby ultra-cool dwarf lying close to the Galactic plane, with spectral characteristics and indices consistent with both early- to mid-L dwarfs and those of the L$/$T transition. We have discussed that its spectral features (especially the pronounced {\em zJH}-band FeH absorption) cannot be fully explained by unresolved binarity when compared with our best fitting composite L and T spectra, although at the same time we cannot completely rule out binarity in 2M1126.

We also find no significant sign of enhanced collision-induced H$_2$ absorption in the {\em H}-band spectrum of 2M1126, which has been shown to suppress the {\em HK}-bands of metal-deficient L sub-dwarfs (\citealt{burg_subd}; \citealt{saumon94}; \citealt*{borysow97}). Given this, and other reasons discussed in \S\ref{subsol}, we suggest that extreme metal-deficiency and/or abnormal gravity are unlikely to be responsible for the discordant features seen in 2M1126, and that 2M1126 is therefore unlikely an L sub-dwarf. However, a {\em K}-band spectrum of 2M1126 is needed to confirm this.

Instead we propose that these characteristics are more suggestive of a single L-T transition object consistent with the condensate cloud clearing model of B02, in which holes in the cloud layers allow flux from warmer, deeper layers still rich in FeH to modify the NIR spectrum. With this interpretation the strength of the FeH absorption seen in 2M1126 implies that the condensate cloud layer has broken-up/thinned at an earlier spectral type across the L-T transition, revealing warmer layers with temperatures consistent with an $\sim$L3 to L5 spectral type (i.e., from the FeH index results). The \JK~colour of 2M1126 fits in well with this scenario, indicating that a substantial outer radiative zone with a large optical depth in the photosphere has not yet developed above these warmer layers. The resurgence of FeH normally seen in T dwarfs peaks at $\sim$T3, with \JK~colours of $\la$0.7, and is at a much lower level than that seen in 2M1126.

As in the case of 2M1126 the blue L dwarfs also show strong FeH and H$_2$O absorption, and a spread of implied spectral type based on their {\em zJHK}-band spectral indices. We suggest that blue L dwarfs could be transitional objects potentially forming part of the natural sequence of objects which undergoing a transition from condensate cloud dominated to clear atmospheres, like those at the canonical L/T boundary, but at slightly higher temperatures (and thus earlier spectral types) due to differences in metallicity. This assumption would lead one to expect a scatter in absolute M{\em $_J$}$/$\JK~diagrams for L-T transition objects at the turn around to bluer \JK~colours, which is seen in empirical data (see fig.1 in B02 and fig.8 of K04). The cloud disruption/break-up hypothesis of B02 combined with this suggestion we present here, offers an explanation for the observable characteristics of blue L dwarfs (and 2M1126), while also associating them with the conventional L-T transition.

Further observations of 2M1126 are highly desirable. Measuring a parallax to establish a reliable distance estimate, and thus absolute {\em J}-band magnitude and temperature, will help to confirm if 2M1126 is a truly single L-T transition object or not. Parallax measurements of the other known blue L dwarfs (currently there no parallaxes available for these objects) will also help to establish where these objects lie in the M{\em $_J/$}\JK~diagram, and if they coincide with the region of partial cloud coverage, compared to the cloud clearing model of B02. Also, the issue of any {\em J}-band brightening, which has recently been comfirmed to be an intrinsic property of early T dwarf atmospheres (from studies of early-T dwarf binaries; B06b), can be addressed for 2M1126. Adaptive optics observations, or HST imaging, are needed to better constrain limits on any companions to 2M1126, which could be detected for separations $\ga0.4$AU at $8.2$pc distant. For separations $\la0.4$AU, radial velocities could address the issue of companions. New optical$/$NIR spectra, and photometric monitoring could highlight variability in the FeH features, CH$_4$ at 1.65 and 2.2$\mu$m, and also in the {\em zJ}-band flux, associated with any changes in the atmospheric dust clouds filling factor/opacity.

\section{Acknowledgements}
This publication makes use of data products from the Two Micron All Sky Survey, which is a joint project of the University of Massachusetts and the Infrared Processing and Analysis Center/California Institute of Technology, funded by the National Aeronautics and Space Administration and the National Science Foundation.
This research has also made use of data obtained from the SuperCOSMOS Science Archive, prepared and hosted by the Wide Field Astronomy Unit, Institute for Astronomy, University of Edinburgh, which is funded by the UK Particle Physics and Astronomy Research Council. This research has made use of the {\sc simbad} database, operated at CDS, Strasbourg, France, and has also benefited from the M-, L-, and T dwarf compendium housed at {\sc dwarfarchives.org} and maintained by Chris Gelino, Davy Kirkpatrick, and Adam Burgasser. The authors would like to thank the reviewers of this paper, Chris Tinney and Adam Burgasser, for their insightful and constructive comments which have contributed to this paper. The authors would also like to thank the staff at the ESO La Silla observatory in Chile.

\bibliography{refs_bibtex}

\bsp
\label{lastpage}

\end{document}